\documentclass[final]{siamltex}

\usepackage{graphicx}

\title{Constructing Generalized Synchronization Manifolds by Manifold Equation}

\author{
	Jie Sun
	\thanks{Department of Mathematics \& Computer Science, Clarkson University,
			Potsdam, NY 13699-5815 ({\tt sunj@clarkson.edu}). This author was supported by the Army Research Office under 51950-MA.}
        \and
        Erik M. Bollt
        \thanks{Department of Mathematics \& Computer Science, Clarkson University,
			Potsdam, NY 13699-5815 ({\tt bolltem@clarkson.edu}).
			This author was supported by the Army Research Office under 51950-MA
			and the National Science Foundation under DMS-0708083 and DMS-0404778.}
	\and
        Takashi Nishikawa
        \thanks{Department of Mathematics \& Computer Science, Clarkson University,
			Potsdam, NY 13699-5815 ({\tt tnishika@clarkson.edu}).}
}

\begin{document}

\maketitle

\begin{abstract}
Full understanding of synchronous behavior in coupled dynamical systems beyond the identical case requires an explicit construction of the generalized synchronization manifold, whether we wish to compare the systems, or to understand their stability. Nonetheless, while synchronization has become an extremely popular topic, the bulk of the research in this area has been focused on the identical case, specifically because its invariant manifold is simply the identity function, and there have yet to be any generally workable methods to compute the generalized synchronization manifolds for non-identical systems. Here, we derive time dependent PDEs whose stationary solution mirrors exactly the generalized synchronization manifold, respecting its stability. 
We introduce a novel method for dealing with subtle issues with boundary conditions
in the numerical scheme to solve the PDE, 
and we develop first order expansions close to the identical case. We give several examples of increasing sophistication, including coupled non-identical Van der Pol oscillators. By using the manifold equation, we also discuss the design of coupling to achieve desired synchronization.\end{abstract}



\pagestyle{myheadings}
\thispagestyle{plain}
\markboth{J. Sun, E. M. Bollt, and T. Nishikawa}{Constructing Generalized Synchronization Manifolds}

\section{Introduction}
Following the fundamental finding in \cite{PECORAprl90} that even chaotic systems can synchronize through coupling, synchronization phenomena in coupled oscillator systems have been studied extensively during the past few years, in the context of dynamical systems (\cite{ALLIGOODchaosbook,OTTchaosbook,RULKOVpre95,ABARBANELpre96,KOCAREVprl96,UCHIDAprl03,ROSENBLUMprl96,ROSENBLUMprl97,MASOLLERprl01}), control theory (\cite{DINGchaos97}), complex networks (\cite{PECORAprl98,ARENASprl06,SORRENTINOpre07}), laser physics (\cite{ MASOLLERprl01,UCHIDAprl03}), etc. 
Besides identical synchronization, studies have been carried out to describe other types of synchronization, such as generalized synchronization~\cite{RULKOVpre95,ABARBANELpre96,KOCAREVprl96,UCHIDAprl03}, phase synchronization \cite{ROSENBLUMprl96,ROSENBLUMprl97}, and anticipated and lag synchronization \cite{ROSENBLUMprl97,MASOLLERprl01}.

The simplest form of synchronization between two oscillators is the complete (or identical) synchronization, meaning that after a transient time, the difference between the two oscillators approaches zero. However, a more generalized form of synchronization can often occur between two non-identical systems.

In the combined state space of such systems, an invariant manifold consisting of the synchronized states, termed \emph{generalized synchronization manifold}, is a central object of study in this context, and its form  (\cite{JOSICprl98}) and stability (\cite{BROWNchaos97,CHUBBijbc01}) are two important properties besides its existence. 
The former tells the relationship between corresponding dynamic variables in the synchronization state, and the latter determines whether a trajectory near the manifold evolves closer or farther away. 
While a great deal of work in synchronization focuses on identical synchronization, primarily because the synchronization manifold 
in this case is described simply by the identity function, much less work (mainly on the detection of existence and stability of generalized synchronization, see \cite{RULKOVpre95,ABARBANELpre96,KOCAREVprl96}) has been carried forth on the ubiquitous generalized synchronization phenomenon, specifically because there has been little work to explicitly construct the corresponding synchronization manifold. One possibility, as reported in \cite{BROWN_PRL98}, is to approximate the manifold using orthogonal functions expansion.

Although one can simply run the system to find a portion of the generalized synchronization manifold that is covered by the attractor of the coupled system, the study of the entire manifold is still necessary and important. One reason is that the transformation from one state variable to the other that defines the manifold contains information on the correlation between the dynamics of the two systems in the whole state space, not just on the attractor. Furthermore, the manifold outside the attractor may embed relevant unstable invariant sets, which can become stable as parameters change. 

Our recent work strives to fill this gap. In this paper we present a PDE (manifold equation) which describes the invariant manifold of generalized synchronization between two coupled oscillators as well as a variational equation regarding the transverse stability of the manifold. 
We also develop a time dependent PDE whose stationary solution is a generalized synchronization manifold,
and present straightforward numerical schemes to solve the equation.  Several examples, such as 1D and 2D coupled oscillators and coupled Van der Pol oscillators are used to illustrate our method of constructing the invariant manifold under generalized synchronization. The design of coupling in coupled system to achieve desired synchronization is also discussed by the use of the manifold equation.

In Section II, we give background regarding generalized synchronization, leading to the manifold equation together with discussion of stability of the manifold and perturbation analysis relative to the identical case. In Section III we show a specific form of the manifold equation regarding coupled oscillator system with almost the same individual dynamics and a time dependent PDE used to solve for the original equation.  In Section IV we show some examples of invariant manifold. In Section V we show how to use the manifold equation to design coupling function between two oscillators to obtain desired form of synchronization.

\section{Equations for synchronization manifold}
For the discussion in this paper, consider the following equations to describe the dynamics of two non-identical oscillators $w_{1}$ and $w_{2}$:
\begin{eqnarray}\label{nonidsys}
\dot{w_{1}}=f(w_{1},w_{2},\mu_1), \nonumber \\
\dot{w_{2}}=g(w_{1},w_{2},\mu_2).
\end{eqnarray}
here $w_{1}\in \Re^m$, $w_{2}\in \Re^m$, $f:\Re^m\times\Re^m\times\Re^1\rightarrow \Re^m$ and $g:\Re^m\times\Re^m\times\Re^1\rightarrow \Re^m$ where $f\in C^{1}$ and $g\in C^{1}$ (both $f$ and $g$ have continuous first order derivative). Note that the coupling functions which enable the two oscillators to synchronize, have been included in the general form of the functions $f$ and $g$. $\mu_1$ and $\mu_2$ are considered as parameters of the functions. 

In this paper, we will focus on perturbations from identical synchronization.  In such a case, we require that $f(w_{1},w_{1},\mu_1)=g(w_{1},w_{1},\mu_1)$, meaning that when the two systems have the same parameters and same value of variables, they have the same individual dynamics, often with no effective coupling between them.

\subsection{Manifold Equation}
Let $(w_{1s},w_{2s})$ denote the synchronization state. The synchronization manifold at synchronization state $w_{2}=\Phi(w_{1})$ is the set $\{(w_{1},w_{2})|w_{2}=\Phi(w_{1})\}$ where $\Phi$ is a time independent transformation. In this paper we will abbreviate this notation and just use the relationship $\Phi$ to indicate the form of this manifold. To find the synchronization manifold of this coupled system, we mean to find some $C^{1}$ function $\Phi$ ($\Phi:\Re^m\rightarrow \Re^m$) such that 
\begin{eqnarray}\label{mffunc}
w_{2s}=\Phi(w_{1s}). 
\end{eqnarray}
Suppose
such $\Phi$ exists.

Then, along the synchronization manifold,
\begin{eqnarray}\label{invmf}
\dot{w_{2s}}=\frac{d\Phi(w_{1s})}{dt}=D\Phi(w_{1s})\cdot\dot{w_{1s}}
\end{eqnarray}

Using Eq.~(\ref{nonidsys}) in Eq.~(\ref{invmf}) 
for the synchronization state, and replacing $w_{1s}$ by $w_{1}$ for convenience, we obtain the following {\it manifold equation}:
\begin{eqnarray}\label{mfeq}
D\Phi(w)\cdot f(w,\Phi(w),\mu_1)=g(w,\Phi(w),\mu_2).
\end{eqnarray}
This is a partial differential equation with unknown function $\Phi=(\Phi_{1},...,\Phi_{m})$ which is of dimension $m$ and each component $\Phi_{i}$ has $m$ variables, and has been used for the study of generalized synchronization \cite{BROWN_PRL98} and center manifold approximation \cite{CARR_BOOK}.

In the case that $\mu_1=\mu_2=\mu$ (coupled identical oscillators), a particular solution of Eq.~(\ref{invmf}) is $\Phi(w)=w$, since $f(w,w,\mu)=g(w,w,\mu)$ which describe the identical manifold under complete synchronization. However, in general cases, it is much harder to find a solution to this PDE. We will leave it to the following to discuss some appropriate approaches in obtaining the solution. 

\subsection{Stability of the Manifold: Variational Equation}
In addition to the form of the manifold, the stability plays another important role in the study of synchronization. To measure the stability of the synchronization manifold $w_{2s}=\Phi(w_{1s})$ obtained by Eq.~(\ref{mfeq}), define
\begin{eqnarray}\label{variation}
\xi \equiv w_{2}-\Phi(w_{1})
\end{eqnarray}
to be the transverse perturbation 
from 
the manifold. Then from Eq.s~(\ref{nonidsys}),~(\ref{invmf}) and ~(\ref{mfeq}), we have
\begin{eqnarray}\label{vart}
\dot{\xi}&=&\dot{w_{2}}-D\Phi(w_{1})\dot{w_{1}} \nonumber\\
         &=&g(w_{1},w_{2},\mu_2)-D\Phi(w_{1})f(w_{1},w_{2},\mu_1) \nonumber\\
        &=&g(w_{1},\Phi(w_{1})+\xi,\mu_2)-g(w_{1},\Phi(w_{1}),\mu_2)+g(w_{1},\Phi(w_{1}),\mu_2)-D\Phi(w_{1})f(w_{1},\Phi(w_{1}),\mu_1) \nonumber\\
        &&+D\Phi(w_{1})f(w_{1},\Phi(w_{1}),\mu_1)-D\Phi(w_{1})f(w_{1},\Phi(w_{1})+\xi,\mu_1).
\end{eqnarray}
The term $g(w_{1},\Phi(w_{1}),\mu_2)-D\Phi(w_{1})f(w_{1},\Phi(w_{1}),\mu_1)=0$ by the manifold Eq.~(\ref{mfeq}), 
so expanding the other terms to the first order in $\xi$, we obtain the {\it local variational equation}:
\begin{eqnarray}\label{vareq}
\dot{\xi}=[D_{w_{2}}g(w_{1},w_{2},\mu_2)-D\Phi(w_{1})D_{w_{2}}f(w_{1},w_{2},\mu_1)]|_{w_{2}=\Phi(w_{1})}\xi.
\end{eqnarray}
This variational equation can be used to study the local stability at the points on the invariant manifold. However, we note that when this equation does not give a uniformly asymptotically stable solution, the asymptotic stability of synchronization may not be correctly predicted from this equation, see examples given in \cite{JOSIC_NONLINEARITY00}. From Eq.~(\ref{vareq}) we again see the importance of the knowledge of the manifold $w_{2s}=\Phi(w_{1s})$. Although there are approaches discussing the stability of the synchronization manifold without knowing explicitly the form of the manifold, it will be more natural and convenient to study the stability directly with the knowledge of the form of the synchronization manifold.

In this paper, we will focus on the form of the invariant manifold. The detailed discussion regarding the stability of the invariant manifold under generalized synchronization will be reported in the future work.

\subsection{Specific Form of Manifold Equation for Perturbed Oscillators} 
Suppose that the individual dynamics for the two oscillators $w_{1}$ and $w_{2}$ in isolation are almost the same, but with a small difference. To be explicit, we mean that in Eq.~\ref{nonidsys} there is a small mismatch in the parameters: $\mu_2=\mu_1+\epsilon$. Then, one would expect the synchronization manifold to be close to the identical synchronization manifold and $\Phi$ to approach the identity function as $\epsilon \to 0$. We empirically expect the synchronization manifold to have the form
\begin{eqnarray}\label{epmf}
w_{2s}=\Phi(w_{1s})=w_{1s}+\epsilon H(w_{1s})+O(\epsilon^2).
\end{eqnarray}
The main interest will be to find $H$ and thus we will have a first order approximation for the synchronization manifold. Following the similar procedure to the derivation of the manifold Eq.~(\ref{mfeq}), after neglecting higher order terms we obtain a PDE for $H$:
\begin{eqnarray}  
DH(w_{1})f(w_{1},w_{1},\mu_1)=&&D_{\mu_2}g(w_1,w_1,\mu_2)|_{\mu_2=\mu_1}\nonumber\\
 &&+D_{w_{2}}[g(w_{1},w_{2},\mu_1)-f(w_{1},w_{2},\mu_1)]|_{w_{2}=w_{1}}H(w_{1}).
\end{eqnarray}

\section{Solving the Manifold Equation in Practice}
In the previous section we derived the general equation for the synchronization manifold of non-identical oscillators as well as the corresponding variational equation for the study of stability of the manifold. We conclude that in order for such transformation $w_{2}=\Phi(w_{1})$ to exist when $t\rightarrow\infty$, (i.e., $\Phi$ is a stationary solution of Eq.~\ref{tdpde} when $t$ is sufficiently large) $\Phi$ has to satisfy Eq.~(\ref{mfeq}). So the problem of finding the invariant manifold reduces to finding the solution of the PDE~(\ref{mfeq}) with appropriate boundary conditions.

However, the boundary condition for Eq.~(\ref{mfeq}) is not easily attainable without actually solving the original ODE system. There may exist more than one synchronization manifolds in general, if none of them is globally stable. Our main interest in this paper is to 
use the perturbed manifold equation (\ref{epmf}) to find the closest one to the identical manifold, which tells us how the form of synchronization changes if the two oscillators are not perfectly the same.

\subsection{A Time Dependent PDE for the Manifold}
In order to develop a workable numerical scheme to solve the manifold equation (\ref{mfeq}), we first derive an evolution equation for a mapping relating the coordinates $w_1$ and $w_2$ of the two systems obeying (\ref{nonidsys}).
The idea is that if we start with a reasonable initial guess for the mapping and evolve it according to the dynamics of the coupled system, the mapping would approach the synchronization manifold, provided that the manifold is asymptotically stable.
Indeed, the synchronization manifold turns out to be a stationary solution of the evolution equation, whose stability mirrors exactly the stability of the synchronized systems.

We assume the existence of a smooth time-dependent mapping from the phase variable $w_{1}$ to $w_{2}$, i.e.,
\begin{eqnarray}
w_{2}=\phi(w_{1},t),
\end{eqnarray}
as an ansatz.
Differentiating this with respect to $t$ and using Eq.~(\ref{nonidsys}), we see that $\phi(w_{1},t)$ satisfies
\begin{eqnarray}\label{tdpde}
\frac{\partial\phi}{\partial t}+\frac{\partial\phi}{\partial w_{1}}f(w_{1},\phi,\mu_1)=g(w_{1},\phi,\mu_2),
\end{eqnarray}
which is a PDE that describes the time evolution of the functional relationship between the states of the two systems.
If we set $\frac{\partial \phi}{\partial t} = 0$, the equation reduces to Eq.~(\ref{mfeq}), so a synchronization manifold, if it exists, is a stationary solution.
Since the evolution equation (\ref{tdpde}) reflects the co-evolution of the coupled systems, we expect a synchronization manifold to be asymptotically stable with respect to Eq.~(\ref{tdpde}), i.e.,
\begin{eqnarray}
\Phi(w_{1})=\lim_{t \rightarrow \infty} \phi(w_{1},t),
\end{eqnarray}
whenever synchronization is asymptotically stable, and the initial condition for Eq.~(\ref{tdpde}) is within the basin of attraction.

If we look for a synchronization manifold of form (\ref{epmf}), then we can write the time-dependent manifold as
\begin{eqnarray}\label{eptdpde}
\phi(w_{1},t)=w_{1}+\epsilon h(w_{1},t)+O(\epsilon^{2})\approx w_{1}+\epsilon h(w_{1},t).
\end{eqnarray}
Neglecting higher order terms in $\epsilon$ we can derive the corresponding time dependent PDE for the function $h(w_{1},t)$:
\begin{eqnarray}\label{pdeh}  
\frac{\partial h}{\partial t}=&&D_{\mu_2}g(w_1,w_1,\mu_2)|_{\mu_2=\mu_1}+D_{w_{2}}[g(w_{1},w_{2},\mu_1)-f(w_{1},w_{2},\mu_1)]|_{w_{2}=w{1}}h\nonumber\\
&&-\frac{\partial h}{\partial w_{1}}f(w_{1},w_{1},\mu_1).
\end{eqnarray}
The stationary state of $h$ gives $H$ corresponding to the synchronization manifold:
\begin{eqnarray}
H(w_{1})=\lim_{t \rightarrow \infty}h(w_{1},t).
\end{eqnarray}

\subsection{An Iterative Scheme to Find the Stationary Manifold}
One approach to solve Eq.~(\ref{pdeh}) is by successive approximation: start with some initial form and evolve the solution according to the equation until the approximation gets close enough to the true solution.
To obtain an iterative scheme to approximate the synchronization manifold, we first discretize $t$ in (\ref{eptdpde}) by considering $t=t_{0},t_{1},...$ where $t_{0}=0$ and $t_{n=1}-t_{n}:=\tau_{n}$. Let us use $h_{n}(w_1)$ to represent $h(w_{1},t_{n})$. Also, for convenience, let
\begin{eqnarray}
b(w_{1})&:=&D_{\mu_2}g(w_1,w_1,\mu_2)|_{\mu_2=\mu_1},\\
B(w_{1})&:=&D_{w_{2}}[g(w_{1},w_{2},\mu_1)-f(w_{1},w_{2},\mu_1)]|_{w_{2}=w_{1}}.
\end{eqnarray}
Then the iteration scheme can be described as
\begin{itemize}
\item Given initial guess $h_{0}$.
\item For $n=1,2,...$, until convergence, do
\begin{eqnarray}\label{iterationh}
h_{n}=h_{n-1}+\tau_{n-1}\Biggl[ b(w_{1})+B(w_{1})h_{n-1}-\frac{\partial h_{n-1}}{\partial w_{1}}f(w_{1},w_{1},\mu_{1})\Biggr].
\end{eqnarray}
\end{itemize}

To analyze the convergence of this scheme, 
let $h^{*}$ denote a fixed point of our iteration operator, i.e.
\begin{eqnarray}
h^{*}=h^{*}+\tau_{n-1}\Biggl[b(w_{1})+B(w_{1})h^{*}-\frac{\partial h^{*}}{\partial w_{1}}f(w_{1},w_{1},\mu_{1})\Biggr].
\end{eqnarray}

Then we have
\begin{eqnarray}
h_{n}-h^{*}&=&h_{n-1}-h^{*}+\tau_{n-1}\Biggl[B(w_{1})(h_{n-1}-h^{*})-\frac{\partial(h_{n-1}-h^{*})}{\partial w_{1}}f(w_{1},w_{1},\mu_{1})\Biggr] \nonumber\\
           &=&(I_{m}+\tau_{n-1}B_{w_{1}})(h_{n-1}-h^{*})-\tau_{n-1}\frac{\partial(h_{n-1}-h^{*})}{\partial w_{1}}f(w_{1},w_{1},\mu_{1})
\end{eqnarray}
so that
\begin{eqnarray}
||h_{n}-h^{*}||\leq||I_{m}+\tau_{n-1}B_{w_{1}}||\cdot||h_{n-1}-h^{*}||+|\tau_{n-1}|\cdot\Bigg|\Bigg|\frac{\partial(h_{n-1}-h^{*})}{\partial w_{1}}\Bigg|\Bigg|\cdot||f(w_{1},w_{1},\mu_{1})||\nonumber\\
\end{eqnarray}
for appropriate matrix and vector norms.

The relevance of each term in this error propagation equation is interpreted as follows. The first term $||I_{m}+\tau_{n-1}B_{w_{1}}||\cdot||h_{n-1}-h^{*}||$ has propagation factor $||I_{m}+\tau_{n-1}B_{w_{1}}||$ smaller than $1$ if synchronization is stable, while the second term $|\tau_{n-1}|\cdot\left|\left|\frac{\partial(h_{n-1}-h^{*})}{\partial w_{1}}\right|\right|\cdot||f(w_{1},w_{1})||$ corresponds to the total variation in space of the error function, which in general is not guaranteed to converge to $0$.

The difficulty of controlling the variation part in the second term above suggests the following adaptive step size scheme:
\begin{eqnarray}\label{adaptiveiterationh} 
h_{n}=h_{n-1}+\tau_{n-1}\left[b(w_{1})+B(w_{1})h_{n-1}-\alpha_{n-1}\frac{\partial h_{n-1}}{\partial w_{1}}f(w_{1},w_{1})\right]
\end{eqnarray}
where $\alpha_{n-1}\in$[0,1] is a control factor that controls the contribution of the total variation to the next iteration: when the total variation is small, choose $\alpha_{n-1}$ close to $1$, and when the total variation is large a small, $\alpha_{n-1}$ will be preferred. A necessary (but not sufficient) condition for the solution to converge is that $\alpha_{n-1} \to 1$ as $n \to \infty$.

\subsection{Spatial Discretization in 2D} 
To be specific, we will demonstrate the 2D discretization in space. The 1D case can be obtained easily from the 2D scheme and for higher dimensions our scheme can be modified to work.

Suppose we have the 2D version of (\ref{iterationh}), and denote $w_{1}=(x,y)'$, and $h(w_{1})=(u(x,y),v(x,y))'$. Also, let 
\begin{eqnarray}
b(w_{1}) = (b_{1}(x,y),b_{2}(x,y))', \nonumber\\ 
f(w_{1},w_{1}) = (f_{1}(x,y),f_{2}(x,y))'
\end{eqnarray}
and $B(w_{1})=$
\[ \left( \begin{array}{cc}
B_{11}(x,y) & B_{12}(x,y) \\
B_{21}(x,y) & B_{22}(x,y) \end{array} \right)\] 
Then the iterations for $u$ and $v$ will be
\begin{eqnarray}\label{iteration2d} 
u_{n} &=& u_{n-1}+\tau_{n-1}\left[b_{1}+(B_{11}u_{n-1}+B_{12}v_{n-1})-\left(\frac{\partial u_{n-1}}{\partial x}f_{1}+\frac{\partial u_{n-1}}{\partial y}f_{2}\right)\right], \nonumber\\
v_{n} &=& v_{n-1}+\tau_{n-1}\left[b_{2}+(B_{21}u_{n-1}+B_{22}v_{n-1})-\left(\frac{\partial v_{n-1}}{\partial x}f_{1}+\frac{\partial v_{n-1}}{\partial y}f_{2}\right)\right],
\end{eqnarray}
given initial guess $u_{0}$ and $v_{0}$.
Suppose we want to solve the equation in a domain $[a,b]\times[c,d]$, we can form a grid on this domain with mesh size $\Delta_{x}$ and $\Delta_{y}$ respectively:
 \begin{eqnarray}
 x_{i} &=& a+i\Delta_{x}, i = 0,1,2,...,N_{x}, \nonumber\\
 y_{j} &=& c+j\Delta_{y}, j = 0,1,2,...,N_{y},
 \end{eqnarray}
 where $N_{x}=\frac{b-a}{\Delta x}$ and $N_{y}=\frac{d-c}{\Delta y}$.
 By using the central difference operators, we have
 \begin{eqnarray}
 \frac{\partial u_{n-1}(i,j)}{\partial x}\approx\frac{u_{n-1}(i+1,j)-u_{n-1}(i-1,j)}{2\Delta_{x}}, \nonumber\\
 \frac{\partial u_{n-1}(i,j)}{\partial y}\approx\frac{u_{n-1}(i,j+1)-u_{n-1}(i,j-1)}{2\Delta_{y}}, \nonumber\\
 \frac{\partial v_{n-1}(i,j)}{\partial x}\approx\frac{v_{n-1}(i+1,j)-v_{n-1}(i-1,j)}{2\Delta_{x}}, \nonumber\\
 \frac{\partial v_{n-1}(i,j)}{\partial y}\approx\frac{v_{n-1}(i,j+1)-v_{n-1}(i,j-1)}{2\Delta_{y}},
 \end{eqnarray}
 the iteration scheme for 2D problem (\ref{iteration2d}) at each grid point $(x_{i},y_{j})$ will be
 \begin{eqnarray}\label{2Ddisit}
 u_{n} &=& u_{n-1}+\tau_{n-1}\Biggl[b_{1}+(B_{11}u_{n-1}+B_{12}v_{n-1}) \nonumber\\
 	      &&-\left(\frac{u_{n-1}(i+1,j)-u_{n-1}(i-1,j)}{2\Delta_{x}}f_{1}+\frac{u_{n-1}(i,j+1)-u_{n-1}(i,j-1)}{2\Delta_{y}}f_{2}\right)\Biggr], \nonumber\\
v_{n} &=& v_{n-1}+\tau_{n-1}\Biggl[b_{2}+(B_{21}u_{n-1}+B_{22}v_{n-1}) \nonumber\\
 	     &&-\left(\frac{v_{n-1}(i+1,j)-v_{n-1}(i-1,j)}{2\Delta_{x}}f_{1}+\frac{v_{n-1}(i,j+1)-v_{n-1}(i,j-1)}{2\Delta_{y}}f_{2}\right)\Biggr].
 \end{eqnarray}
Here the functions without arguments are evaluated at grid points $(x_{i},y_{j})$.

\subsection{On Boundary Condition and Computation}
There is one issue for implementing 
the iteration (\ref{2Ddisit}): 
the values on the boundary of the domain is difficult to compute in general.
One approach to avoid this issue is to use dynamically shrinking domains. Start by assigning values at all points in a initial domain that is much larger than the domain of interest. 
Then the iteration is defined properly at all interior points, but not on the boundary, so we make the domain smaller by discarding all boundary points for the next iteration.
Repeating this process, we produce a sequence of domains of decreasing size on which successively better approximation of the synchronization manifold is obtained.
By taking the initial domain to be large enough so that the iteration converges before the domain becomes smaller than the original domain of interest, we obtain an approximate solution on the entire domain of interest.

Intuitively, this method works by choosing the boundary of the computational domain to dynamically change with the propagation wave front of the effect of the boundary of the initial domain.  Since the solution at a given fixed point in the domain will not be affected by the choice of values on the boundary of the initial domain until this wave front reaches the point, any point that remains in our shrinking domain should not be affected by the choice boundary condition on the initial domain. The influence of the unknown boundary condition of the outer domain becomes progressively less important as time progresses and the inner domain shrinks.

\section{Examples of Invariant Manifold}
As we discussed in the previous sections, the manifold equation (\ref{mfeq}) is important because it allows further study of the synchronization between coupled oscillators. In this section we show several examples of constructing a generalized synchronization manifold.
In particular, we consider examples in which the dynamics of the two oscillators are different but only with minor difference. A motivation for the study of this special case is that in any real dynamical systems no oscillator is physically perfect in the sense that the parameters or even the structure we use to mold these oscillators bear some small error.

\subsection{A Coupled 1D Oscillators}
One-dimensional examples provide an excellent starting point of investigation for studying invariant manifolds for higher-dimensional systems.
Consider a simple case where
\begin{eqnarray}\label{speq}
\dot{x}=1, \quad x \in \Re.
\end{eqnarray}
We couple this oscillator with another one, but with some perturbation function on it so that the two oscillators are not identical:
\begin{eqnarray}\label{simple1D}
\dot{x}&=&1, \nonumber\\
\dot{y}&=&1+(x-y)+\epsilon(p(x)+p'(x)).
\end{eqnarray}
where $p$ is any smooth function and $p'$ denotes its derivative. 
Here the $(x-y)$ term represents the one-way coupling from $x$ to $y$ and the $\epsilon(p(x)+p'(x))$ term represents some small perturbation on the $y$ oscillator. Note that when $\epsilon=0$ the synchronization manifold is $y=x$, which is stable.  
This oscillator is designed to have $\Phi(x)=p(x)$ as a stable invariant manifold, which can be checked by introducing an appropriate change of coordinates. This type of oscillator always results in an intrinsic relationship between $x$ and $y$ subcomponent, given by $y=p(x)$.

For example, if let $p(x)=\sin(x)$, then we have
\begin{eqnarray}\label{simple1D2}
\dot{x}&=&1, \nonumber\\
\dot{y}&=&1+(x-y)+\epsilon(\sin(x)+\cos(x)).
\end{eqnarray}
Now for $\epsilon\neq0$, the invariant manifold $\Phi(x)$ is determined by the manifold equation (\ref{mfeq}), which yields
\begin{eqnarray}\label{simple1Dmfeq}
\frac{d\Phi(x)}{dx}=-\Phi(x)+(1+x)+\epsilon(\sin(x)+\cos(x)).
\end{eqnarray}
This equation is a simple ODE that can be solved analytically using the multiplication factor method. The solution is
\begin{eqnarray}\label{simple1Dmf}
\Phi(x)=x + \epsilon \sin(x) + Ce^{-x}
\end{eqnarray}
where $C$ is a constant that 
is determined by the boundary condition. 
In this family of invariant manifolds, the only one that corresponds to the perturbation of the identical synchronization ($\Phi(x) \to x$ when $\epsilon \to 0$) is obtained when $C=0$: 
\begin{eqnarray}\label{simple1Dmf2}
\Phi(x)=x+\epsilon \sin(x).
\end{eqnarray}
The stability of the manifold can be determined by (\ref{vareq}), in this case we obtain
\begin{eqnarray}\label{simple1Dvar}
\dot{\xi}=-\xi,
\end{eqnarray}
so for $t\rightarrow \infty$, the invariant manifold is exponentially stable.

\begin{figure}[htbp]
	\includegraphics[height=5.00in,width=6.00in]{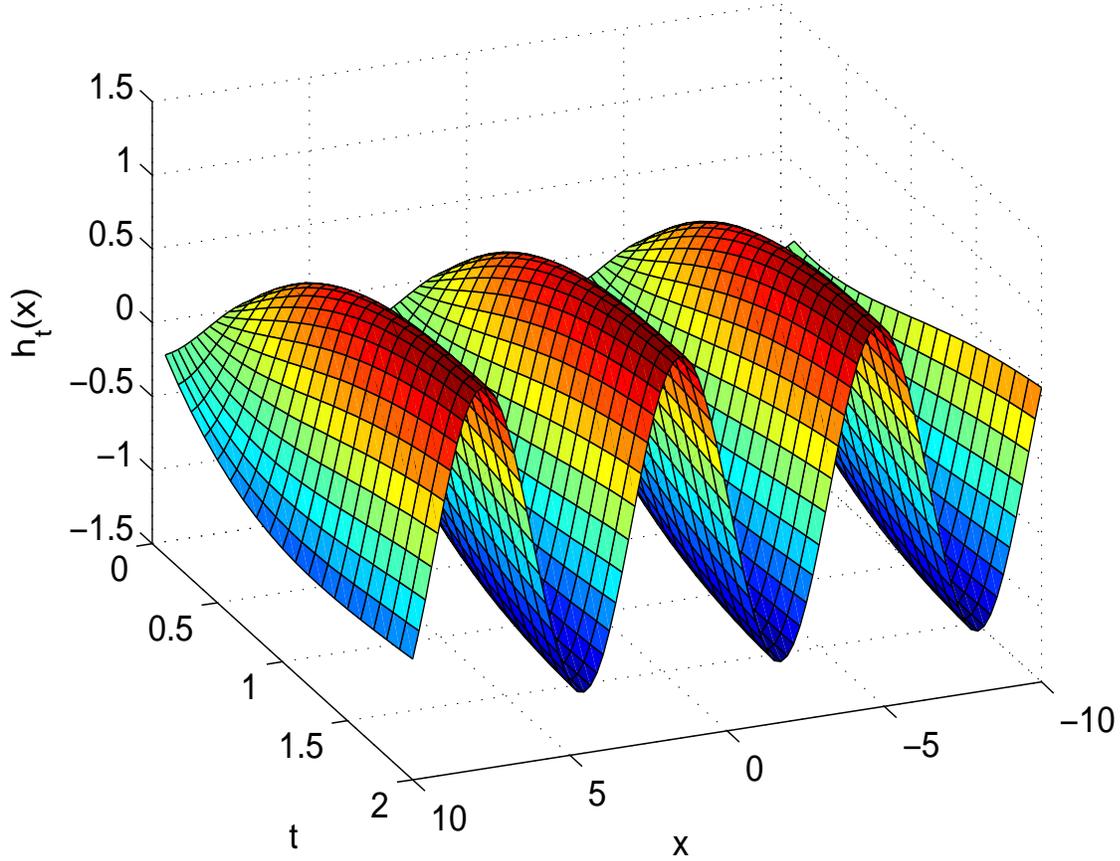}
\caption{Successive approximation of the manifold form $H(x)$ corresponding to the system (\ref{simple1D}), here $h_{t}(x)$ corresponds to successive functions obtained from (\ref{iterationh}), with step size $\tau=0.1$. The approximation tends to the true form $H(x)=\sin(x)$.} 
\label{pmanifold}
\end{figure}

The manifold can also be obtained by using the iteration scheme we described in the previous section: suppose the form of the manifold $\Phi(x)=x+\epsilon H(x)$. To find $H$ we look for the corresponding stationary solution of the PDE described by (\ref{pdeh}). Following the iteration scheme (\ref{iterationh}), we start with $h_{0}=0$ and choose $\tau_{n}=\tau=0.1$, and obtain the following sequence of functions: 
\begin{eqnarray}
h_{0} &=& 0, \nonumber\\
h_{1} &=& 0.1\cdot \sin(x)+0.1\cdot \cos(x), \nonumber\\
h_{2} &=& 0.2\cdot \sin(x)+0.18\cdot \cos(x), \nonumber\\
h_{10} &=& 0.8340089600\cdot \sin(x)+0.3315041568\cdot \cos(x), \nonumber\\
&&... \nonumber\\
h_{100} &=& 0.9999965626\cdot \sin(x)-0.00004893563553\cdot \cos(x), \nonumber
\end{eqnarray}
As we can see, this sequence of iteration functions converge to $\sin(x)$, which is the true form that appears in the manifold. We plot the first $20$ iterates in Fig.~\ref{pmanifold} to visualize the convergence of iterations.

\subsection{Coupled 2D Oscillators }
We now consider two coupled oscillators and study the invariant manifold that describes the long-term relationship between them. 
More specifically, consider a unidirectionally coupled system made up of two different oscillators:
\begin{eqnarray}\label{pqoscillator}
\dot{x_{1}}&=&1, \nonumber\\
\dot{y_{1}}&=&1+(x_{1}-y_{1})+\epsilon_{1}(p_{1}(x_{1})+p_{1}'(x_{1})), \nonumber\\
\dot{x_{2}}&=&1+k(x_{1}-x_{2}), \nonumber\\
\dot{y_{2}}&=&1+(x_{2}-y_{2})+\epsilon_{2}(p_{2}(x_{2})+p_{2}'(x_{2})).
\end{eqnarray}

Here the first oscillator is sending signal to the second through the $x$ component, where $k$ is the coupling strength. For large enough $k$ we expect the corresponding $x$ components to synchronize identically, while the form of synchronization of the $y$ component can depend 
on the choice of $\epsilon$ and $p$. However, note that when $x$ synchronizes, the coupling term becomes small and we have two individual oscillators which we know how to solve. For $t\rightarrow\infty$ we expect to have the following relations to hold, at least in the approximate sense:
\begin{eqnarray}\label{x2y1y2}
x_{2}&=&x_{1}, \nonumber\\
y_{1}&=&x_{1}+\epsilon_{1}p_{1}(x_{1}), \nonumber\\
y_{2}&=&x_{2}+\epsilon_{2}p_{2}(x_{2}).
\end{eqnarray}
Let $\epsilon:=\epsilon_{2}-\epsilon_{1}$ measure the order of mismatch of the two oscillators, and we look for the relationship $y_{2}=y_{1}+\epsilon H(x_{1},y_{1})$. Plugging Eq.~(\ref{x2y1y2}) into this equation, we obtain
\begin{eqnarray}
\epsilon H(x_{1},y_{1})=\epsilon_{2}p_{2}(x_{1})-\epsilon_{1}p_{1}(x_{1}). 
\end{eqnarray}
Thus, we expect the invariant manifold between these two oscillators to be given by:
\begin{eqnarray}
x_{2}&=&x_{1}, \nonumber\\
y_{2}&=&y_{1}+\epsilon_{2}p_{2}(x_{1})-\epsilon_{1}p_{1}(x_{1}).
\end{eqnarray}
Indeed, one can show that these equations do give an exponentially stable synchronization manifold by introducing an appropriate change of variables.

As an example, let us consider $p_{1}(x)=p_{2}(x)=\sin(x)$, but with $\epsilon_{1}\neq\epsilon_{2}$.
Setting $\epsilon=\epsilon_{2}-\epsilon_{1}$, the invariant manifold is given by
\begin{eqnarray}
x_{2}&=&x_{1}, \nonumber\\
y_{2}&=&y_{1}+\epsilon \sin(x_{1}).
\end{eqnarray}
To check the iteration scheme in this case, let
\begin{eqnarray}
x_{2}&=&x_{1}+\epsilon u(x_{1},y_{1}), \nonumber\\
y_{2}&=&y_{1}+\epsilon v(x_{1},y_{1})
\end{eqnarray}
be the form of invariant manifold
and apply the iteration scheme (\ref{iterationh}) with $h(x_1, y_1) = (u(x_1,y_1), v(x_1,y_1))'$. 

Starting with the initial functions $u_{0}=v_{0}=0$, 
we find that
\begin{eqnarray}
u_{0} = 0, v_{0} &=& 0, \nonumber\\
u_{1} = 0, v_{1} &=& 0.1\cdot \sin(x)+0.1\cdot \cos(x), \nonumber\\
u_{2} = 0, v_{2} &=& 0.2\cdot \sin(x)+0.18\cdot \cos(x), \nonumber\\
u_{10} = 0, v_{10} &=& 0.8340089600\cdot \sin(x)+0.3315041568\cdot \cos(x), \nonumber\\
&&... \nonumber\\
u_{100} = 0, v_{100} &=& 0.9999965626\cdot \sin(x)-0.00004893563553\cdot \cos(x), \nonumber
\end{eqnarray}

and $u_{n}$ converges to $0$ while $v_{n}$ converges to $\sin(x_{1})$, as expected. 

\subsection{Coupled Non-Identical Van der Pol Oscillators}
\begin{figure}[htbp]
	\includegraphics[height=3.50in,width=5.50in]{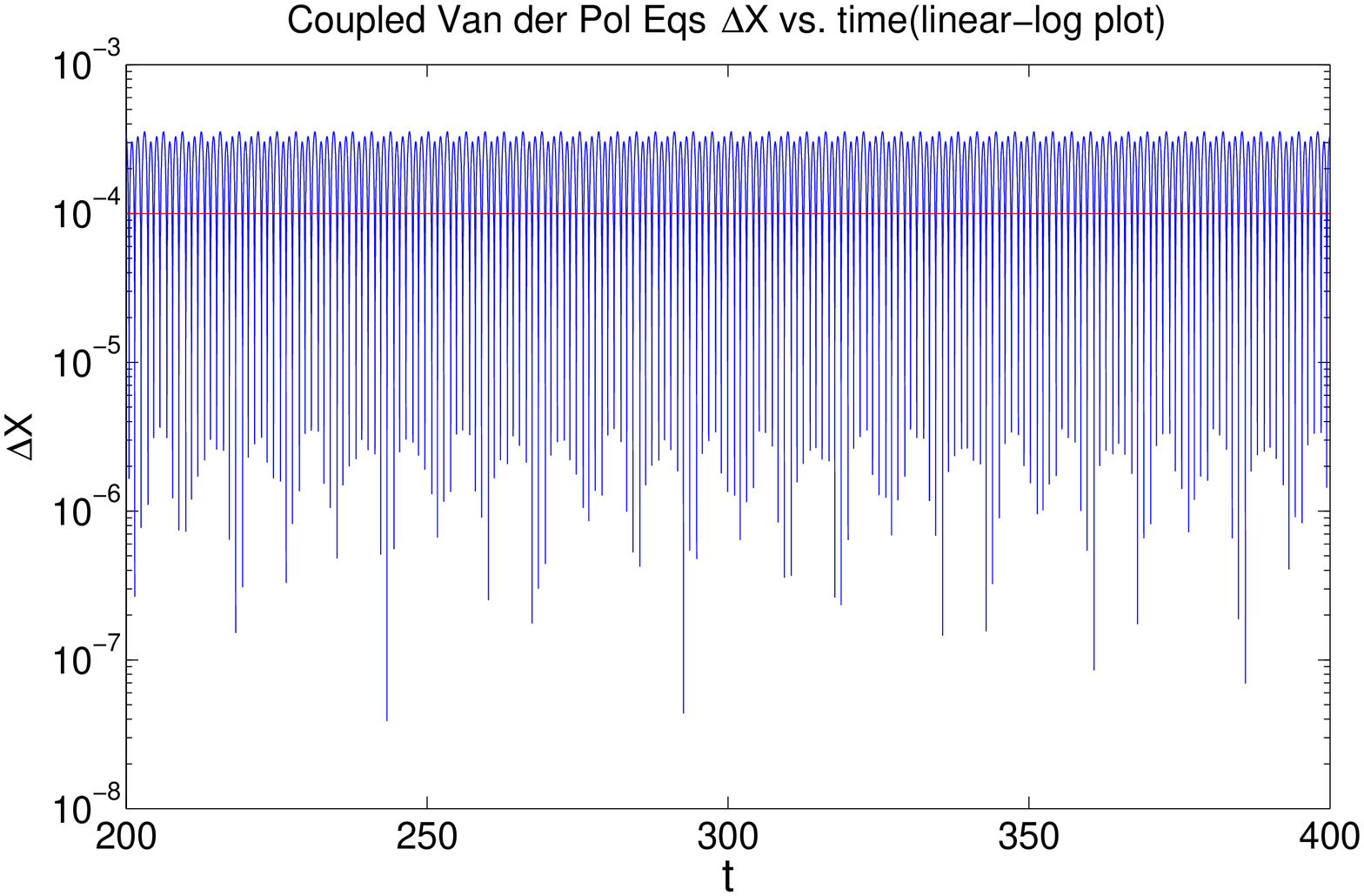}
	\vspace{0.50in}
	\includegraphics[height=3.50in,width=5.50in]{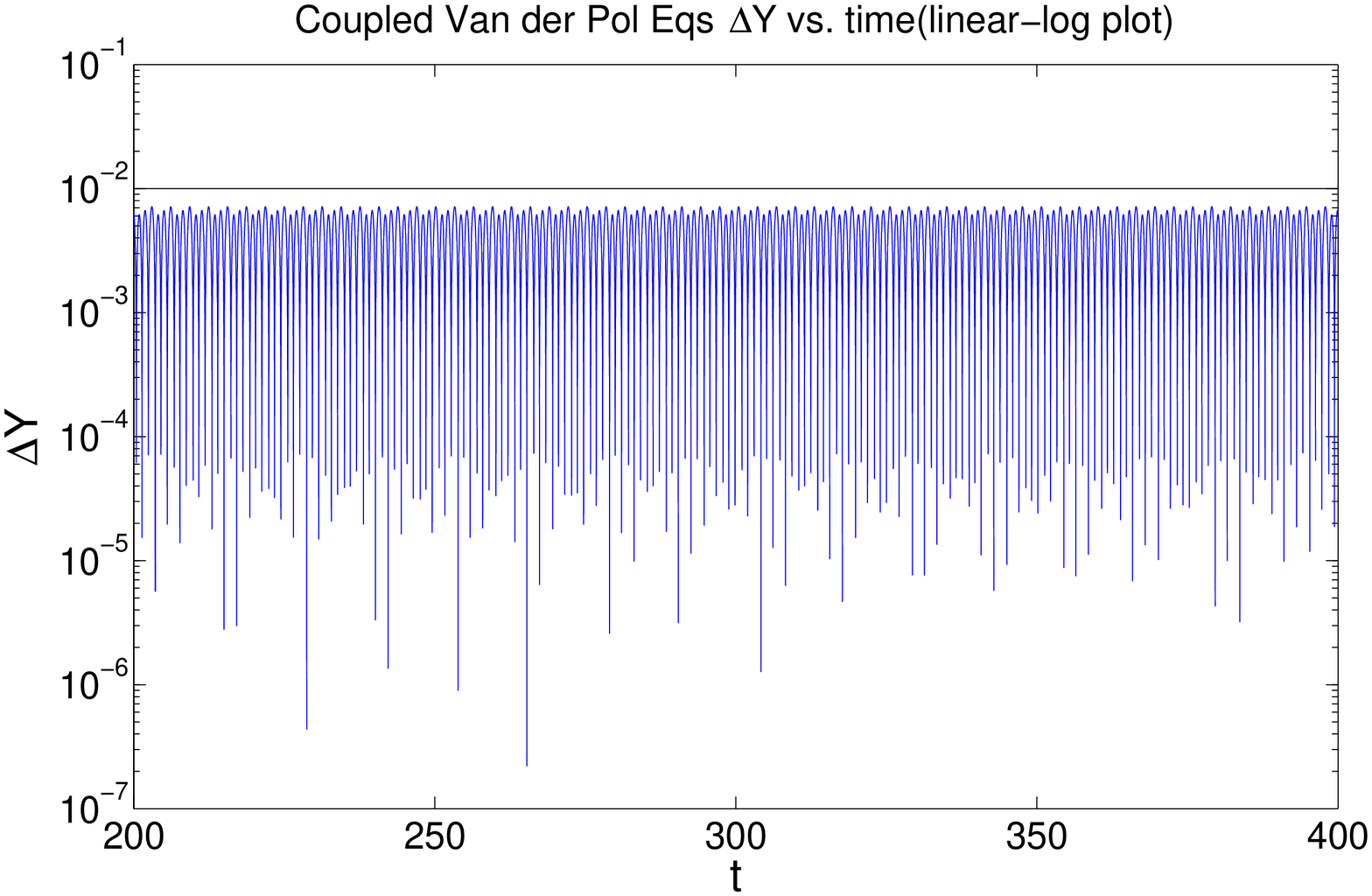}
\caption{Deviation from the identical manifold. The time series comes from the system (\ref{cpdvanderpol1}) starting from $\bigl(x_{1}(0),y_{1}(0),x_{2}(0),y_{2}(0)\bigr)=(1.5,1.5,1.5006,1.5107)$. Here in the two panels we plot the interval $t\in[200,400]$. The curve in the upper panel shows the difference $\Delta X(t) := |x_{1}(t)-x_{2}(t)|$ and in the lower panel we plot $\Delta Y(t) := |y_{1}(t)-y_{2}(t)|$. Red line in the upper panel shows the value $\epsilon^{2}=10^{-4}$ and the black line in the lower panel indicates $\epsilon=10^{-2}$.}
\label{VDPxxyy}
\end{figure}

The equation of Van der Pol oscillator can be written as a first order system:
\begin{eqnarray}\label{singlevdp}
\dot{x}&=&y, \nonumber\\
\dot{y}&=&-x+\mu(1-x^{2})y
\end{eqnarray}
where $x$ represents position, $y$ velocity, and $\mu$ is the strength of the nonlinear damping. For $\mu \ll 1$, there exists a stable limit cycle of radius $2$. Now consider two of these oscillators, with slightly different values of the parameter $\mu$, and coupled through the $x$ component from the first to the second. The equations describing this system are:
\begin{eqnarray}\label{cpdvanderpol1}
\dot{x_{1}}&=&y_{1}, \nonumber\\
\dot{y_{1}}&=&-x_{1}+\mu(1-x_{1}^{2})y_{1}, \nonumber\\
\dot{x_{2}}&=&y_{2}+k(x_{1}-x_{2}), \nonumber\\
\dot{y_{2}}&=&-x_{2}+(\mu+\epsilon)(1-x_{2}^{2})y_{2}.
\end{eqnarray}
Here in the second oscillator the $k(x_{1}-x_{2})$ term represents the one-way coupling from the first oscillator, and $\epsilon \ll \mu$ is the mismatch in the damping strength. Since here we have two different individual systems coupled together, the identical manifold is not invariant under this dynamics. However, for large enough $k$, when $t\rightarrow\infty$ we numerically observe that the difference of $x_{1}$ and $x_{2}$ is $O(\epsilon^2)$ while the difference of $y_{1}$ and $y_{2}$ is $O(\epsilon)$. In Fig.~\ref{VDPxxyy} we show an example of the deviation of trajectory from the identical manifold, with parameters chosen to be $\mu=0.1$, $\epsilon=0.01$ and $k=20$. Generalized synchronization appears instead of complete synchronization in this case. 

\begin{figure}[htbp]
	\includegraphics[height=7.00in,width=6.00in]{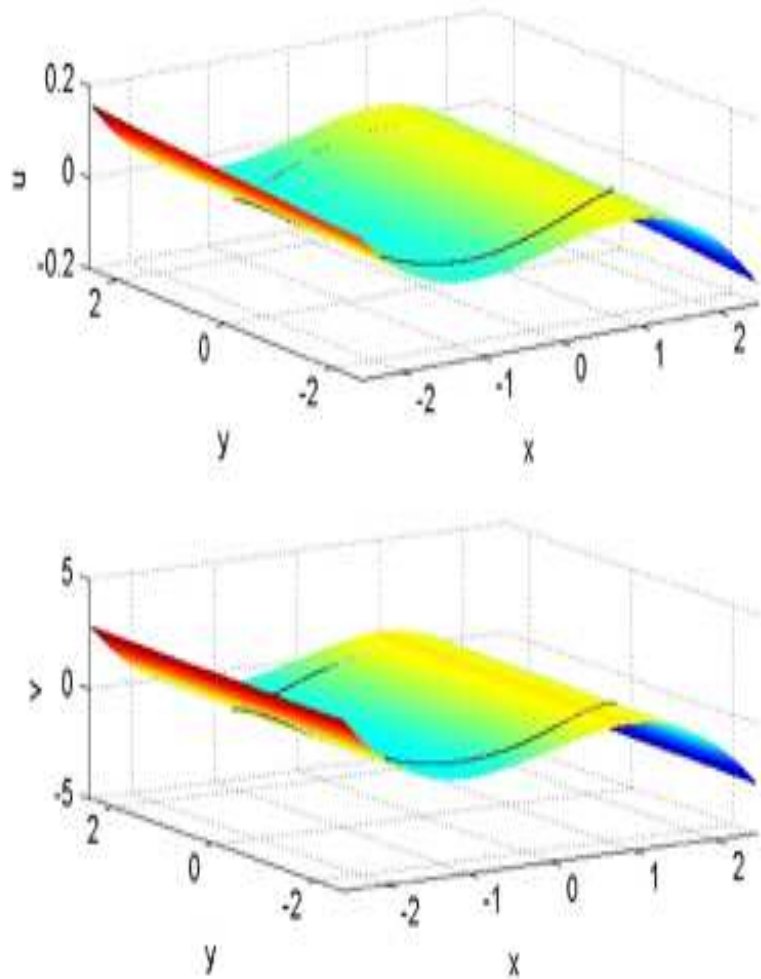}
\caption{Numerical solution of the time dependent PDE corresponding to the system (\ref{cpdvanderpol1}). In the upper panel we plot $u$, and the black curve landing on $u$ is obtained from plotting time series $\left(x_{1}(t),y_{1}(t),\frac{x_{2}(t)-x_{1}(t)}{\epsilon}\right)$ obtained from original ODE. Similarly, in the bottom panel we plot $v$ as well as the time series $\left(x_{1}(t),y_{1}(t),\frac{y_{2}(t)-y_{1}(t)}{\epsilon}\right)$.} 
\label{VDPmanifold}
\end{figure}

To demonstrate the effectiveness of our numerical scheme, use the 2D iteration scheme (\ref{iteration2d}), start with initial guess close to the theoretical prediction $u(x,y)=0,v(x,y)=x-\frac{1}{3}x^{3}$ (see Appendix for derivation), and compute the iterations starting on the domain: $[-5,5]\times[-5,5]$ with mesh size $0.005$ and time spacing $\tau=0.0005$. We terminate when the domain shrinks down to $[-2.5,2.5]\times[-2.5,2.5]$ and take the solution at that time to be the approximated stationary solution. In Fig.~\ref{VDPmanifold} we plot the solutions $u$ and $v$, with a typical trajectory obtained from the original ODE system. 

\begin{figure}[htbp]
	\includegraphics[height=3.00in,width=4.00in]{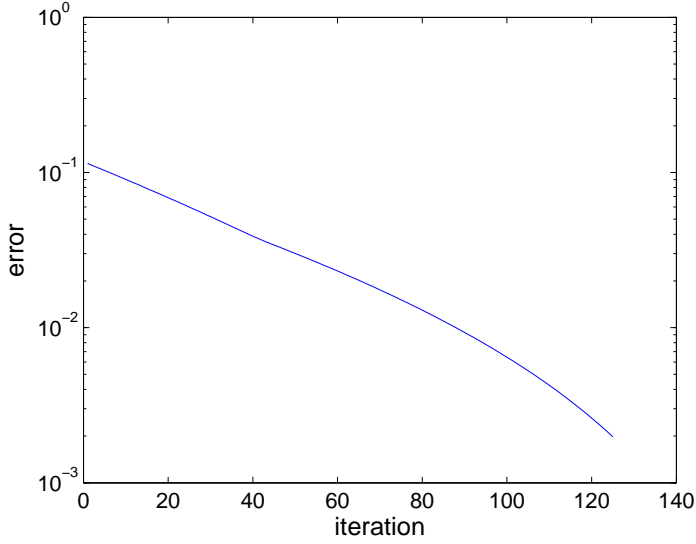}
\caption{Successive error $e_n$ in the 2D iteration scheme for the time dependent PDE of the non-identical coupled Van der Pol system. Here the error is measured by the infinite norm.}
\label{VDPerror}
\end{figure}

\begin{figure}[htbp]
	\includegraphics[height=3.50in,width=5.00in]{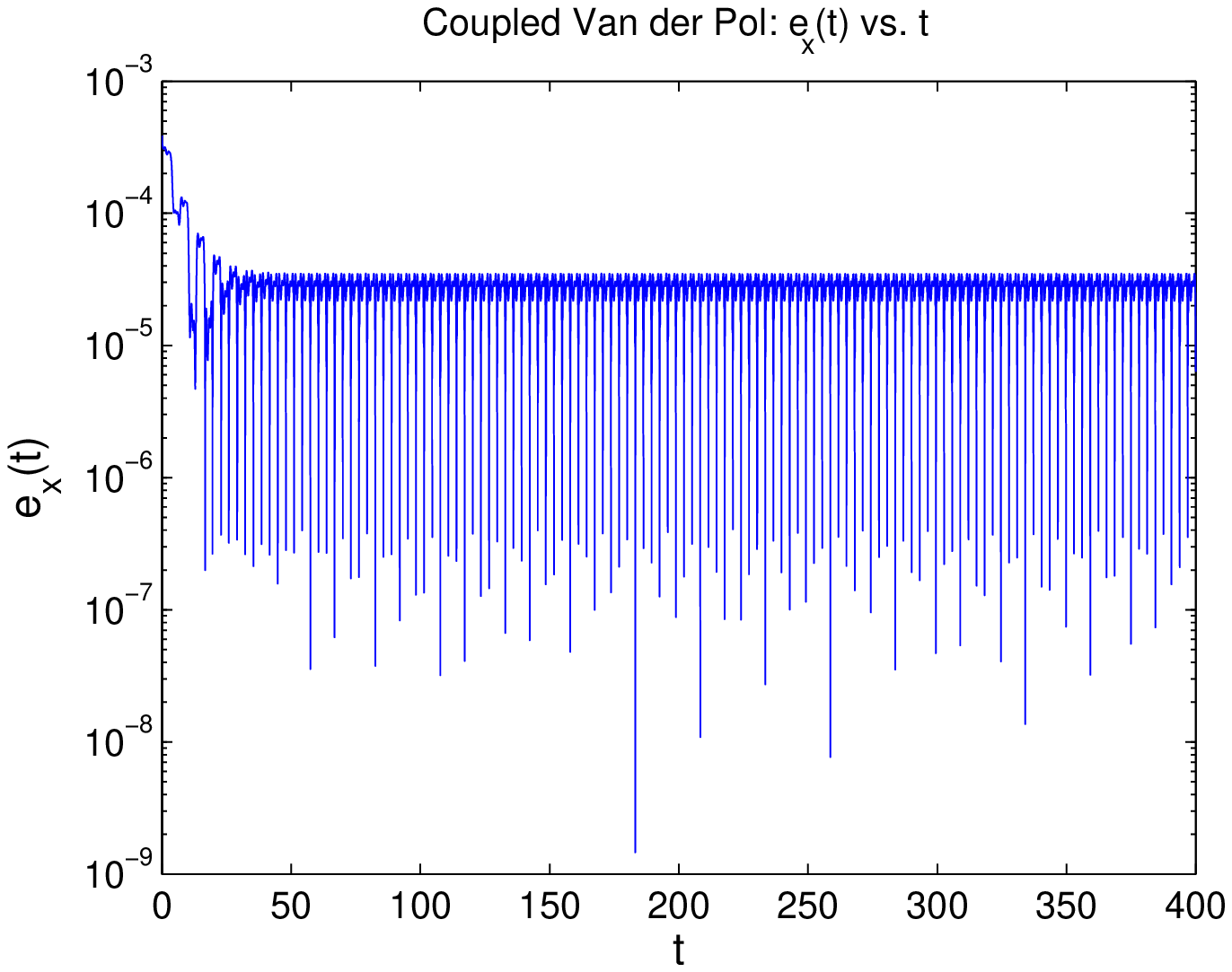}
	\vspace{0.50in}
	\includegraphics[height=3.50in,width=5.00in]{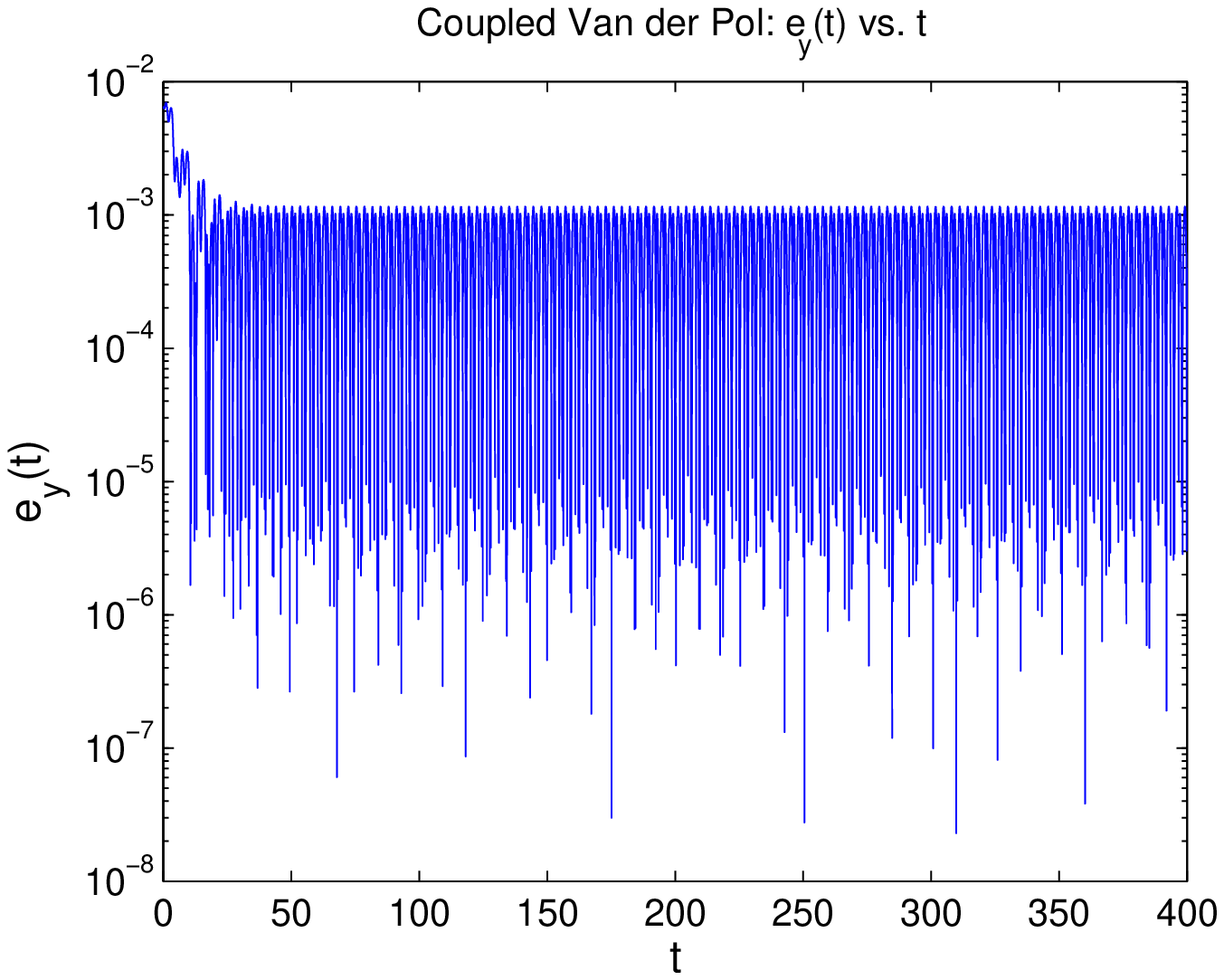}
\caption{Distance of a typical trajectory to the manifold obtained by numerical scheme. Here $e_{x}(t):=|x_{2}(t)-x_{1}(t)-\epsilon u(x_{1}(t),y_{1}(t))|$ and $e_{y}(t):=|y_{2}(t)-y_{1}(t)-\epsilon v(x_{1}(t),y_{1}(t))|$ where $u$ and $v$ are solutions by numerical scheme on grid points and for points in between grid points we use spline interpolation to obtain its corresponding value. Here we see that $e_{x}\sim o(\epsilon)$ and $e_{y}\sim o(\epsilon)$, suggesting that the first order approximation in $\epsilon$ ($\epsilon=0.01$) is achieved.}
\label{VDPmanifold3}
\end{figure}

Figure~\ref{VDPerror} shows the successive error $e_n = \max(||u_n - u_{n-1}||_{\infty}, ||v_n - v_{n-1}||_{\infty})$ measured by the infinite norm.  We see exponential decrease in this error, which indicates that the iterative scheme is converging.  In Fig.~\ref{VDPmanifold3} we plot the distance of a typical trajectory to the manifold obtained by this scheme, which suggests that the computed manifold indeed gives a good first-order approximation for the synchronization manifold. 

\section{Design Coupling to Satisfy Desired Synchronization}
In many real applications it is useful to have the knowledge of how to design specific coupled system so that the resulting behavior satisfy certain purposes. Mathematically we are seeking for a method to design the coupling between two general oscillators so that the resulting invariant manifold will have the prescribed form. In this section we discuss how to achieve the above goal by making use of the manifold equation we derived earlier. 

Assume that without coupling we have two separate oscillators described by the following equations:
\begin{eqnarray}\label{dfos}
\dot{w_{1}}&=&F(w_{1}), \nonumber \\
\dot{w_{2}}&=&G(w_{2}).
\end{eqnarray}
where $w_{1}\in \Re^m$, $w_{2}\in \Re^m$, $F:\Re^m\rightarrow \Re^m$ and $G:\Re^m\rightarrow \Re^m$ where $F\in C^{1}$ and $G\in C^{1}$. Now suppose we want $w_{1}$ and $w_{2}$ to synchronize, and furthermore, the synchronization manifold is prescribed as $w_{2}=\Phi(w_{1})$, the question is how to couple the two oscillators so that the resulting invariant manifold will have the desired form and be stable.

\subsection{Form of Coupling Function}
We first solve for the coupling function so that the corresponding coupled system will have the invariant manifold satisfying the prescribed form. For convenience, we use a one-way coupling, so that the system becomes
\begin{eqnarray}\label{cpddfos}
\dot{w_{1}}&=&F(w_{1}), \nonumber \\
\dot{w_{2}}&=&G(w_{2})+K(w_{1},w_{2}).
\end{eqnarray}
where $K:\Re^m\times \Re^m\rightarrow \Re^m$ represents the coupling from $x$ to $y$. In order not to make the question too broad, we assume that the coupling function has the form
\begin{eqnarray}\label{cplfunc}
K(w_{1},w_{2})=\sigma(L(w_{1})-H(w_{2}))
\end{eqnarray}
where $L:\Re^m\rightarrow \Re^m$ and $H:\Re^m\rightarrow \Re^m$ both $\in C^{1}$. $\sigma$ is a scalar which measures the strength of coupling, often called the {\it coupling coefficient}, or {\it coupling strength}. Then in order that the synchronization manifold of the corresponding coupled system 
\begin{eqnarray}\label{cpddfos2}
\dot{w_{1}}&=&F(w_{1}), \nonumber \\
\dot{w_{2}}&=&G(w_{2})+\sigma(L(w_{1})-H(w_{2}))
\end{eqnarray}
has the desired form $w_{2}=\Phi(w_{1})$, by equation (\ref{mfeq}), we have:
\begin{eqnarray}\label{cplfunc2}
D\Phi(w_{1})F(w_{1})=G(\Phi(w_{1}))+\sigma(L(w_{1})-H(\Phi(w_{1})).
\end{eqnarray}
Note here that we have two functions $L$ and $H$ to choose while the manifold only gives one constraint, we need to fix one of the coupling forms. Suppose we are given the form of $H(w_{2})$, and want to find $L(w_{1})$. Solving for $L(w_{1})$ yields
\begin{eqnarray}\label{cplfunc3}
L(w_{1})=\frac{1}{\sigma}[D\Phi(w_{1})\cdot F(w_{1})-G(\Phi(w_{1}))]+H(\Phi(w_{1})).
\end{eqnarray}
So for any given form of the synchronization manifold $w_{2}=\Phi(w_{1})$, we are able to design the corresponding coupling between the two oscillators. 

Notice that for identical oscillators ($F=G$), if we want $\Phi(w_{1})=w_{1}$ (identical synchronization), then
\begin{eqnarray}\label{idjacobi}
D\Phi(w_{1})=I_{m},
\end{eqnarray}
is the $m\times m$ identity matrix so the first term in Eq.~(\ref{cplfunc3}) disappears and the second term is just $H(\Phi(w_{1}))=H(w_{1})$, so we have
\begin{eqnarray}\label{cplfunc4}
L(w_{1})=H(w_{1}),
\end{eqnarray}
meaning that for the coupled system (\ref{cpddfos2}) to have the identical synchronization manifold, the coupling functions $L$ and $H$ should have the same form, as expected.

\subsection{Adjusting Coupling Coefficient to Obtain Stability}
In Eq.~(\ref{cplfunc3}) we have the form of the coupling function which guarantees that the coupled system described by Eq.~(\ref{cpddfos2}) have the invariant manifold $w_{2}=\Phi(w_{1})$. The corresponding variantional equation along the manifold can be obatined using Eq.~(\ref{vareq}), 
which becomes
\begin{eqnarray}\label{vareq2}
\dot{\xi}=D[G(w_{2})-\sigma H(w_{2})]|_{w_{2}=\Phi(w_{1})}\xi.
\end{eqnarray}
Stability can be obtained by altering the value of $\sigma$, the coupling strength, if it is possible to have stable synchronization. Furthermore, with this variational equation as well as the form of the invariant manifold (known in advance), we are able to analyze the stability of not only the whole manifold but also individual points or subsets on the invariant manifold.

\subsection{Examples of Application}
One case of the design of coupling form appears in linear systems with perturbations. Suppose we have two uncoupled linear systems:
\begin{eqnarray}\label{lnsys}
\dot{w_{1}}&=&Aw_{1}, \nonumber\\
\dot{w_{2}}&=&Bw_{2}
\end{eqnarray}
where $A$ and $B$ are linear operators (matrices). It is known that if $A=B$ then by linear diffusive coupling the two oscillators will synchronize and the synchronization is stable for large enough coupling strength. Now in the more general situation $A$ and $B$ can be different, either because the true systems are not made perfectly exact, or other reasons. But we still want to couple them so that they achieve identical synchronization. The analysis in the previous sections offer us a way to design the coupling form.

Again use one way coupling and assume the form of coupling function on $w_{2}$. The coupled system becomes
\begin{eqnarray}\label{cpdlnsys} 
\dot{w_{1}}&=&Aw_{1}, \nonumber\\
\dot{w_{2}}&=&Bw_{2}+\sigma(L(w_{1})-Hw_{2})
\end{eqnarray}
where $H$ is a coupling matrix (known in advance) and $L$ is the unknown coupling function of $w_{1}$ that needs to be determined. By Eq.~(\ref{cpddfos2}) and (\ref{cplfunc3}), we find that:
\begin{eqnarray}\label{lncpl}
L(w_{1})=\left(\frac{A-B}{\sigma}+H\right)w_{1}.
\end{eqnarray}
Here we see that only when $A=B$ we can allow $L(w_{1})=H(w_{1})$ and yield identical synchronization, otherwise the coupling form of $x$ and $y$ shall be different to compensate for the difference of the original dynamics on $x$ and $y$ respectively. 
The stability of the manifold is described by
\begin{eqnarray}\label{lnvar}
\dot{\xi}=(B-\sigma H)\xi
\end{eqnarray}
following Eq.~(\ref{vareq2}). In order for the synchronization manifold to be stable, we need
\begin{eqnarray}\label{lnlmd}
\rho(B-\sigma H)<1,
\end{eqnarray} 
where $\rho(Q)$ is the spectral radius of the matrix $Q$.

Eq.s~(\ref{lncpl}), (\ref{lnvar}) and (\ref{lnlmd}) gives us a way to easily manipulate the coupling functions in the non-identical linear systems to obtain stable identical synchronization. With minor modifications we can obtain similar results for two-way coupling, and with a few more modifications as well as more analysis we shall be able to control more complicated systems.

\section{Discussion and Conclusion}
In this paper we 
have developed systematic methods for explicit construction of generalized synchronization manifolds in systems of coupled non-identical oscillators by means of the manifold equation, a PDE that must be satisfied by the manifold, and the associated variational equation describing its stability.

Although the manifold equation gives necessary condition to determine the invariant manifold, it is not sufficient, mainly due to the fact that boundary conditions for this PDE are not easily attainable. 
To obtain a solution of the manifold equation without knowing explicitly the boundary conditions, we have proposed a time dependent PDE whose stably stationary solution is a solution of the manifold equation, and developed iteration schemes to 
solve it both symbolically and numerically.
Several examples of constructing synchronization manifolds have been given for systems of nearly identical oscillators  (considered as a perturbed version of identical oscillators, which correspond to complete synchronization) with unidirectional coupling, where we obtained the first order approximation of the perturbed manifold.
A general technique for using the manifold equation to design the coupling for the purpose of controlling the form and stability of synchronization has also been discussed and illustrated using the simple case of linear systems. 

Our numerical algorithms used to obtain the stationary solution of the time dependent PDE is analogous to the Euler method in a functional space, 
whose convergence is not necessarily global.  Thus, finding a systematic method to choose an appropriate initial condition, as well as a detailed convergence analysis for the scheme, is an important problem that must be addressed in the future. As there exists a variety of different schemes to solve PDEs in the literature, exploring and comparing with other numerical schemes for solving the manifold equation is another important topic of future research.

\section*{Acknowledgements}
J.S. and E.M.B have been supported for this work by the Army Research Office grant 51950-MA. E.M.B. has been further supported by the National Science Foundation under DMS-0708083 and DMS-0404778. We would like to thank Naratip Santitissadeekorn for discussion in the numerical scheme part.

\section*{Appendix}
In the coupld Van der Pol system (\ref{cpdvanderpol1}), in order to obtain the first order approximation of the synchronization manifold in $\epsilon$, we neglect higher order terms in $\epsilon$ and write
\begin{eqnarray}
x_{2}&=&x_{1}+O(\epsilon^2)\approx x_{1}, \nonumber\\
y_{2}&=&y_{1}+\epsilon V(x_{1},y_{1})+O(\epsilon^2)\approx y_{1}+\epsilon V(x_{1},y_{1}).
\end{eqnarray}
Thus, using Eq.~(\ref{cpdvanderpol1}), we have
\begin{eqnarray}
\dot{y_{2}}-\dot{y_{1}}&=&\epsilon(1-x_{1}^{2})y_{1}+O(\epsilon^{2}). \nonumber\\
\end{eqnarray}
We also have, for $\mu \ll 1$,
\begin{eqnarray}
\dot{y_{2}}-\dot{y_{1}}&=&\epsilon\left(\frac{\partial V}{\partial x_{1}}\dot{x_{1}}+\frac{\partial V}{\partial y_{1}}\dot{y_{1}}\right)+O(\epsilon^{2}) \nonumber\\
                                        &=&\epsilon\left\{y_{1}\frac{\partial V}{\partial x_{1}}+[-x_{1}+\mu(1-x_{1}^{2})y_{1}]\frac{\partial V}{\partial y_{1}}\right\}+O(\epsilon^{2}) \nonumber\\
                                        &=&\epsilon\left(y_{1}\frac{\partial V}{\partial x_{1}}-x_{1}\frac{\partial V}{\partial y_{1}}\right)+O(\epsilon^2).
\end{eqnarray}

Replacing $x_{1}$, $y_{1}$ with $x$ and $y$ for convenience and equating the first-order $\epsilon$ terms in the above two equations, we obtain the following equation:
\begin{eqnarray}\label{VDPpde}
y\frac{\partial V}{\partial x}-x\frac{\partial V}{\partial y}=(1-x^{2})y
\end{eqnarray} 
This equation has a particular solution:
\begin{eqnarray}
V(x,y)=x-\frac{1}{3}x^{3}
\end{eqnarray}

Since the difference between $x$ components in this case is already $O(\epsilon^2)$, we take the identical manifold $x_{2}=x_{1}$ as the first order approximation, while for the $y$ component we use the form obtained above: $y_{2}=y_{1}+\epsilon V(x_{1},y_{1})$.

\end{document}